\documentclass{aa}
\usepackage{graphicx}

\usepackage{txfonts}
\begin{document}

     \title{Basis function expansions for galactic dynamics:  Spherical versus cylindrical coordinates}
      \author{Yougang Wang;
        \inst{1}\fnmsep\thanks{wangyg@bao.ac.cn}
         E.~Athanassoula;
         \inst{2}
         Shude Mao
         \inst{3,1}
        } 
     \institute{Key Laboratory of Computational Astrophysics, National Astronomical Observatories,Chinese Academy of Sciences, Beijing, 100012 China
          \and  
          Aix Marseille Univ, CNRS, CNES, LAM, Marseille, Franc
          \and
         Department of Astronomy, Tsinghua University, Beijing 100084, China
       }
%\date{Received Apr 15, 2020; accepted Apr 30, 2020} 

\abstract
{}
{The orbital structure of galaxies is strongly influenced by the accuracy of the force calculation during orbit integration. We explore the accuracy of force calculations for two expansion methods and determine which one is preferable for orbit integration.}    
%methods heading (mandatory)
{We specifically compare two methods, one was introduced by Hernquist \& Ostriker (HO), which uses a spherical coordinate system and was built specifically
for the Hernquist model, and the other by Vasiliev \& Athanassoula (CylSP) has a cylindrical coordinate system. Our comparisons include the Dehnen profile, its triaxial extension (of which the Hernquist profile is a special case) and a multicomponent 
system including a bar and disk density distributions for both analytical models and $N$-body realizations.}
{For  the generalized Dehnen density, the CylSP method is more accurate than the HO method for nearly all inner power-law indices and shapes  
 at all radii. For $N$-body realizations of the Dehnen model, or snapshots of an $N$-body simulation, the CylSP method is more accurate than the HO method in the central region for the oblate, prolate, and triaxial Hernquist profiles if the particle number is more than $5\times 10^5$. For snapshots of the Hernquist models with spherical shape, the HO method is preferred.
For the Ferrers bar model, the force from the CylSP method is more accurate than the HO method. 
The CPU time required for the initialization of the HO method  is significantly shorter than that for the CylSP method, while the HO method costs subsequently much more CPU time than the CylSP method if the input corresponds to particle positions.
From surface of section analyses, we find that the HO method creates more chaotic orbits than the CylSP method in the bar model.
This could be understood to be due to a spurious peak in the central region when the force is calculated with the HO expansion.}
{For an analytical model, the CylSP method with an inner cutoff radius of interpolation $R_{\rm min}$ as calculated by the AGAMA software, is preferred due to its accuracy.  
For snapshots or $N$-body realizations not including a disk or a bar component, a detailed comparison between these two methods is needed if a density model other than the Dehnen model is used. 
For multicomponent systems, including a disk and a bar, the CylSP method is preferable.}


%
%-------------------------------------------------------------------------------------
\titlerunning{Basis function expansion methods}
\authorrunning{Wang et al.}
\maketitle{}

\section{introduction}
Stellar orbits and their families are the fundamental building blocks for the dynamics of galaxies. 
They are also critical in constructing dynamical models of galaxies using the Schwarzschild method ~\citep[e.g.][]{1979ApJ...232..236S,1997ApJ...488..702R,2004MNRAS.353..391T,2006A&A...445..513V,2008MNRAS.385..647V,2008ApJ...677.1033W,2009MNRAS.396..109W,2013MNRAS.433.3173B,2017ApJ...844..130W,2018MNRAS.473.3000Z,2019MNRAS.484.1166M,2019MNRAS.486.4753J} or the made-to-measure method ~\citep[e.g.][]{1996MNRAS.282..223S,2007MNRAS.376...71D,2009MNRAS.395...76D,2012MNRAS.421.2580L,2013MNRAS.432.3062H,2013MNRAS.428.3478L,2014ApJ...792...59Z,2016RAA....16..189L,2017MNRAS.470.1233P,2018MNRAS.473.2288B}.
Moreover, orbits can be used to explain tidal streams ~\citep[e.g.][]{2016ApJ...824..104P,2017NatAs...1..633P}, the kinematics in the Solar neighborhood ~\citep[e.g.][]{1998MNRAS.298..387D,2001A&A...373..511F}, the origin of hypervelocity stars ~\citep[e.g.][]{2014ApJ...785L..23Z,2014ApJ...793..122K,2019MNRAS.483.2007E}, the flow of gas in bars ~\citep{1992MNRAS.259..328A,1992MNRAS.259..345A} and the formation and structure of bars, including their boxy and peanut bulges ~\citep[e.g.][]{1997ApJ...483..731P,2002MNRAS.337..578P,2003MNRAS.342...69P}.
% the bar-induced evolution of dark matter cusps ~\citep[e.g.][]{2005MNRAS.363..991H}. 
%{\bf All these results depend on the accuracy of the orbit calculations in  analytical analyses or simulations.} 

The studies mentioned above often require orbit integrations. For simple analytic potentials, the accuracy of orbit integration is high. However, most of these models are not sufficiently realistic for galactic studies. In more realistic models, for example for
the bar in the Milky Way, the potential and force cannot be given directly in an analytical form, even for an analytic density profile. Therefore, methods for calculating the potential and force are needed even for analytical models.

A direct way to obtain the orbit properties is by using $N$-body simulations. We note that $N$-body simulations have successfully reproduced many properties of galaxies, but it is difficult to control the accuracy of the force calculation in a simulation.  The advantage of  $N$-body simulations is that the potential models are self-consistent; therefore, the orbit properties can be studied in a frozen  simulation potential as long as it varies slowly. One accurate method to calculate the force in such a potential is to sum the force contribution for each particle; however, this method is not realistic if the particle number is large. An alternative is to calculate density-potential pairs using basis-function expansion techniques ~\citep[e.g.][]{2016MNRAS.459.3349R}. The Poisson's equation can be solved in 
spherical coordinates ~\citep[e.g.][]{1973Ap&SS..23...55C,1992ApJ...386..375H,1996MNRAS.283..149Z,2009MNRAS.393.1459R,2011MNRAS.416.2697L,2018MNRAS.476.2092L} or in cylindrical coordinates ~\citep[e.g.][]{1972Ap&SS..16..101C,1995MNRAS.277.1341K,1996ApJ...465...91E,1998MNRAS.294..429D,1999ApJ...527...86C}.

 For each coordinate system, there are many different basis  functions.  One such expansion technique was proposed by \citealt{1992ApJ...386..375H} (hereafter HO92). Hereafter we refer to this method as the HO method. This method has been widely used for galactic bars ~\citep[e.g.][]{1996MNRAS.283..149Z,2012MNRAS.427.1429W,2016ApJ...824..104P,2017NatAs...1..633P}  and for $N$-body potentials ~\citep[e.g.][]{2002ApJ...580..627W,2012MNRAS.422.1863B,2013MNRAS.435.3437W,2014ApJ...792...98M,2017MNRAS.466.3876Z,2019MNRAS.483.3048W}. For the expansion in cylinder coordinates, ~\cite{2015MNRAS.450.2842V} have improved the speed and accuracy of the force calculation by using a spline interpolation technique. We refer to their method as the CylSP method.   

Each of these expansion methods has its advantages and disadvantages, and some recent studies already gave comparisons between the spherical-harmonic expansion method and the CylSP method ~\citep{2015MNRAS.450.2842V,2019MNRAS.482.1525V}. There are many parameters in each method, and the accuracy of each method depends on the values chosen for these parameters, as well as on the density distribution and the particle number. In this paper, we give here a more detailed comparison between the HO and CylSP methods. The accuracy of the force calculation can affect the orbit shape, and the resulting orbit families in the systems~\citep[e.g.][]{2006CeMDA..96..129C}. In this paper we also study how the expansion methods affect the orbit families. The outline of the paper is as follows. In Section 2 we describe the two expansion methods for the potential and accelerations. In Section 3 we compare their accuracy for a triaxial extension of the Dehnen models (Dehnen 1993). In Section 4 we present our comparisons for a 
multicomponent model which includes a Ferrers bar. In Section 5 we present the summary and discussions.

\section{Potential and accelerations}
Given a model density distribution, the potential can be calculated  using Poisson's equation. Then the accelerations of the model can be obtained from the derivatives of the potential. For complicated density distributions, the potential and density can be expanded on a set of simple orthogonal basis-function in spherical coordinates          
\begin{equation}
\rho(r,\theta,\phi)=\sum_{l=0}^{l_{max}}\sum_{m=0}^{l}\sum_{n=0}^{n_{max}}A_{nlm}\rho_{nl}Y_{lm}(\theta,\phi),
\end{equation}  
  
\begin{equation}
\Phi(r,\theta,\phi)=\sum_{l=0}^{l_{max}}\sum_{m=0}^{l}\sum_{n=0}^{n_{max}}B_{nlm}\Phi_{nl}Y_{lm}(\theta,\phi),
\end{equation}  
where $n$ is the radial expansion parameter, and $l$ and $m$ are the usual angular parameters.  There are many basis functions and different expansion formats in the spherical coordinates ~\citep[e.g.][]{1973Ap&SS..23...55C,1992ApJ...386..375H,1996MNRAS.283..149Z,1999AJ....117..629W,2012MNRAS.427.1429W,2013MNRAS.434.3174V}.
Here we focus on the HO method. The potential and accelerations can be written as Equations (3.13), (3.21), (3.22) and (3.23) in HO92, respectively. In the HO method, the expansion coefficients are determined by  three-dimensional integrations (See Eq. 3.14 in HO92). For a given system, the expansion coefficients are calculated only once, and the potential and force in any arbitrary 
position can be obtained using these coefficients.

For a nearly disk system, the density and potential pairs are naturally expanded in cylindrical coordinates ($R$, $z$, $\phi$). The density and potential can be represented by the Fourier series  
\begin{equation}
\rho(R,z,\phi)=\sum_{m=0}^{m_{\rm max}}\rho_m(R,z)\exp(im\phi),
\end{equation}  
\begin{equation}
\Phi(R,z,\phi)=\sum_{m=0}^{m_{\rm max}}\Phi_m(R,z)\exp(im\phi),
\end{equation}
and each term in the expansion is  given by
\begin{equation}
\rho_m(R,z)=\frac{1}{2\pi}\int_0^{2\pi}{\rm d}\phi\rho(R,z,\phi)\exp(-im\phi),
\end{equation}

\begin{equation}\label{eq:phi_m}
\Phi_m(R,z)=-G\int_{-\infty}^{+\infty}\int_{0}^{\infty}{\rm d}z^{\prime}{\rm d}R^{\prime}2\pi R^{\prime}\rho_m(R^{\prime},z^{\prime})\ \Xi.
\end{equation}
and 
\begin{equation}
\Xi(R,z,R^{\prime},z^{\prime})=\int_0^{\infty}{\rm d}k J_m(kR)J_m(kR^{\prime}\exp(-k \left|z-z^{\prime}\right|),
\end{equation}
where $m_{\rm max}$ is the truncated number of azimuthal Fourier harmonic terms. For an axisymmetric model, only the $m=0$ term is needed, while $m_{\rm max}\lesssim 12$ is enough for typical asymmetric models~\citep{2019MNRAS.482.1525V}. We note that $J_m$ is an $mth-$order Bessel function of the first kind,  and can be expressed analytically by the half-integer degree Legendre function of the second kind $Q_{m-1/2}$~\citep{1965tisp.book.....G,1999ApJ...527...86C,2005MNRAS.363..991H}
\begin{equation}
\Xi=\frac{1}{\pi\sqrt{RR^{\prime}}}Q_{m-1/2}\bigg(\frac{R^2+{R^{\prime}}^2+(z-z^{\prime})^2}{2RR^{\prime}} \bigg)\ ,    {\rm if}\ R, R^{\prime}>0, 
\end{equation}

\begin{equation}
\Xi=[R^2+{R^{\prime}}^2+(z-z^{\prime})^2]^{-1/2}\delta_{m0} ,\ {\rm if}\ R=0\ {\rm or}\ R^{\prime}=0.
\end{equation}
Equation~\ref{eq:phi_m} shows that we know that the potential can be obtained by a two-dimensional integration, and this integration depends on the $(R,z)$ coordinates of the system. 
Therefore, the calculation is time consuming if the input model is analytical. In order to increase the speed for the calculations of potential and forces, ~\cite{2015MNRAS.450.2842V} improved this expansion using a spline interpolation. Therefore, the potential and forces can only be obtained only in some fixed grids ($N_R\times N_z$)  within a fixed region $R\in[R_{\rm min}, R_{\rm max}]$, $z\in[z_{\rm min}, z_{\rm max}]$,  and then the potential and forces in any position can be derived by the spline interpolation. There are two advantages in using the spline interpolation. First, the speed for the orbit integration can be improved by several orders of magnitude. Second, the unphysical noise for the force can be smoothed if the input model is
a snapshot of an $N$-body simulation.

\section{ Generalized Dehnen models}
\subsection{General density}
The  density models we consider here are triaxial and their spherical analogs were first discussed by ~\cite{1993MNRAS.265..250D}. The density model is ~\citep{1996ApJ...460..136M}  
\begin{equation}\label{eq:dehnen}
\rho(m)=\frac{(3-\gamma)M}{4abc}m^{-\gamma}(1+m)^{-(4-\gamma)},\ \ \  0\leq\gamma<3,
\end{equation}
with
\begin{equation}
m^2=\frac{x^2}{a^2}+\frac{y^2}{b^2}+\frac{z^2}{c^2},\ \   a\geq b\geq c.
\end{equation}
Above, $M$ is the total mass of the model and $\gamma$ is the inner density profile index. We refer to this kind of density distribution as the generalized Dehnen models or simply as Dehnen models.  Specifically, the $\gamma=1$ case is the same as the ~\cite{1990ApJ...356..359H} model. The iso-density contours are ellipsoids with axis ratios of $a$:$b$:$c$. We adopt four combinations of  $a$:$b$:$c$ = 1:1:1, 1:1:0.2, 1:0.2:0.2, and 1:0.79:0.5, corresponding to spherical, oblate, prolate, and triaxial models, respectively.
~\cite{1996ApJ...460..136M} analytically reduced the triple integrals of the potential and force calculations to a single integral each. This is of course a considerable improvement as the corresponding calculations are much easier to perform numerically and, moreover,  are much more accurate. We thus use them here, labeling them as ``true" values. For a spherical model, the potential and forces have analytical expressions (See Equations 4 and 6 in ~\citealt{1996ApJ...460..136M}). We have checked the accuracy for the true values obtained using one-dimensional integrals relative to analytic values 
in the spherical model, and we found that they match each other within a relative accuracy of $10^{-15}$ for the model with $\gamma=1$ and better than $10^{-8}$ for the other values of $\gamma$.
It is important to notice that the $x$-axis, is the major axis while the $z$-axis is the minor axis.  Henceforth, we use units such that the total mass $M$, the major axis length $a$, and the gravitational constant $G$ are equal to unity.   

\subsection{Comparisons for analytic potentials}
\begin{figure*}
\includegraphics[angle=0, width=180mm]{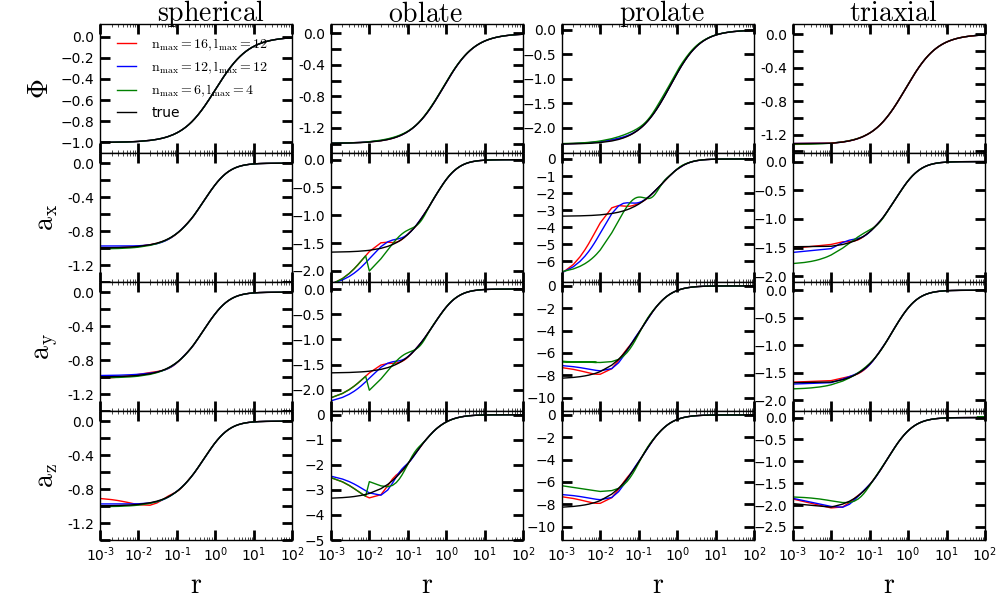}
\caption{Potential and forces for the Hernquist models. From left to right, the input density model is spherical, oblate, prolate, and triaxial, respectively. In each panel, the black line is the result for the model given by Merritt  \& Fridman (1996) . The red, blue, and green lines are the results for the HO method with ($n_{\rm max}$, $l_{\rm max}$)=(16,12), (12,12), and (6,4) , respectively. From top to bottom, the radius $r$ was measured along the $x-$, $x-$, $y-$, and $z-$ axes, respectively. The $x$-axis is the major axis and the $z$-axis is the minor axis of the density distribution.}
\label{fig:Hern_HO_ana}
\end{figure*}

In Figure~\ref{fig:Hern_HO_ana}, we show the potential and accelerations of the Hernquist models for the true values and the results from the HO method. It is seen that the HO method can approximate the potential well for a spherical system. For oblate and prolate systems, the  HO method does not  provide accurate forces in the central region. Naturally,  larger values of $n_{\rm max}$ and $l_{\rm max}$ give more accurate expansions than smaller ones for most models and in most regions. For a spherical system, we find that large values of $n_{\rm max}$ and $l_{\rm max}$ generate larger errors in the force along the minor axis $(a_z)$ in the central region than those only using the lower order
terms of coefficients, this is reasonable and consistent with that in HO92. In the HO method, the zeroth order terms in the expansion of the density and potential are the Hernquist distribution and its corresponding potential. Therefore, if only the zeroth terms are adopted in the spherical model, then the errors in both the forces and potential are zero. In order to give a consistent comparison for different shapes, we also kept the high-order terms in the expansion.  
Also, for the triaxial system, if $n_{\rm max}=16$ and $l_{\rm max}=12$, then the HO method can give accurate expansions.

\begin{figure*}
\includegraphics[angle=0, width=180mm]{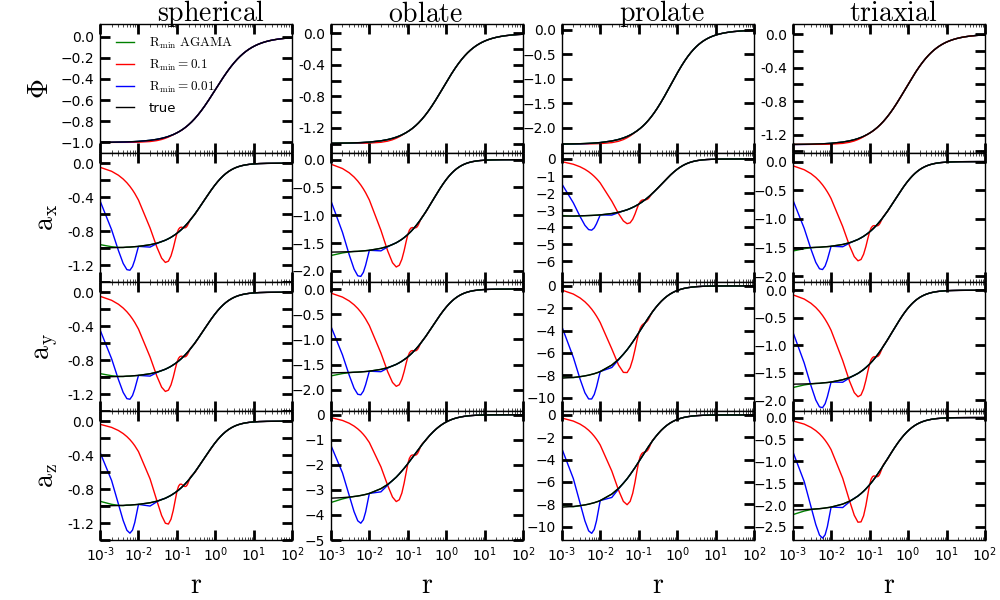}
\caption{Same as in Figure ~\ref{fig:Hern_HO_ana}, but for the results of the CylSP method. In each panel, the red, and blue lines are the results of the CylSP method with $R_{\rm min}=0.1$, and 0.01, respectively, while the green lines correspond to the value calculated  by the AGAMA software (see Sect. 3.2). The black line is the analytical result given by Merritt  \& Fridman (1996). Here $N_{\rm z}=N_{\rm R}=40$, $z_{\rm min}=R_{\rm min}$, and $R_{\rm max}=z_{\rm max}=100$, $m_{\rm max}=12$.}
\label{fig:Hern_agama_ana}
\end{figure*}

In Figure~\ref{fig:Hern_agama_ana}, we show the potentials and accelerations of the Hernquist models, comparing the true values and the results from the CylSP method. Here we consider different values of $R_{\rm min}$ in this method.
In the AGAMA software implementation ~\citep{2019MNRAS.482.1525V} of the  CylSP method that we are using,  the code  automatically  
chooses the $R_{\rm min}$ value and in the case of our models gives a few times $10^{-4}$.
 It is clear that this method can reproduce the potential well for spherical, oblate, prolate, and triaxial systems. For the forces from the CylSP method, the value of  $R_{\rm min}$ that is given by the software can give accurate values in all cases except
in the central-most region ($r<2\times10^{-3}$) for the spherical, prolate, and  triaxial models. For the forces from $R_{\rm min}=0.01$ and $R_{\rm min}=0.1$, obvious wiggles appear in the central region ($r<0.1$).

We also consider different values of $m_{max}$ and $N_R$, $N_z$ for the CylSP method. It should be noted that $N_R=40$ can give good  accuracy as shown in ~\cite{2015MNRAS.450.2842V}. 
We then searched for the optimum $m_{\rm max}$ value and found that there is hardly any improvement beyond $m_{\rm max}=12$, while the necessary CPU time is considerably increased. 
For $m_{\rm max}\geqslant 12$, the accuracy for the force and potential is almost unchanged even when $m_{\rm max}$ is further increased together with CPU time. Therefore, we adopt $N_R=N_z=40$ and $m_{\rm max}=12$ in the CylSP method in this paper unless stated otherwise.

\begin{figure*}
\includegraphics[angle=0, width=180mm]{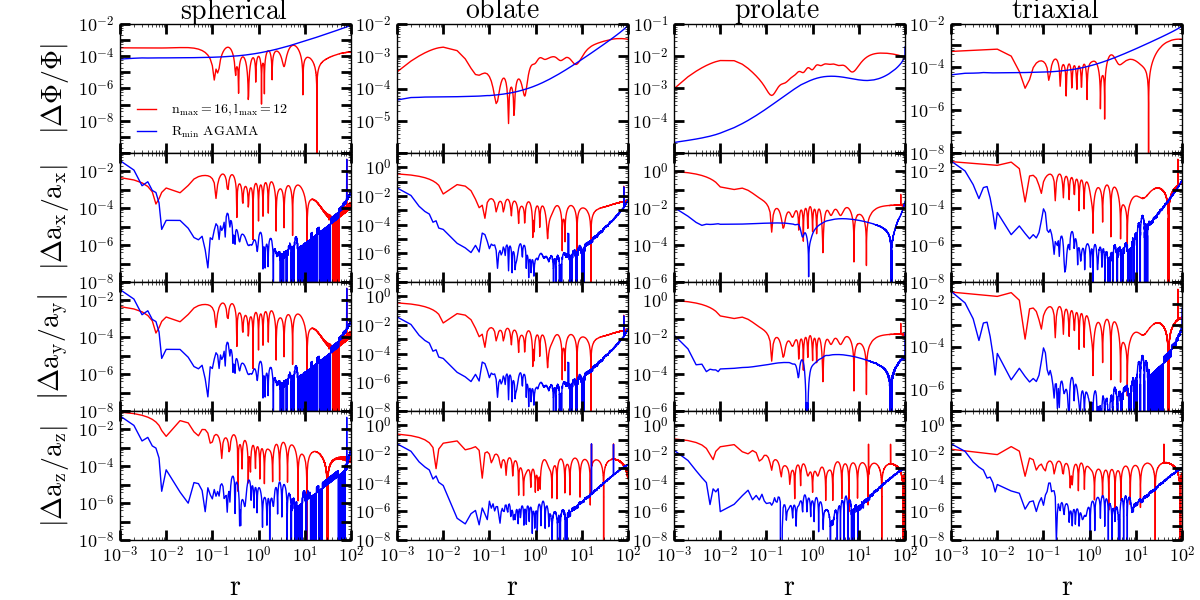}
\caption{Relative error in the potential and forces for both the HO (red lines) and CylSP (blue lines) methods for the analytical Hernquist models. For the HO method, ($n_{\rm max}$, $l_{\rm max}$)=(16,12). For the CylSP method, $R_{\rm min}$ is given by the AGAMA software, $N_R=40$, $N_z=40$, $z_{\rm min}=0$, $R_{\rm max}=z_{\rm max}=100$.}
\label{fig:Hern_agama_comp_ana}
\end{figure*}

In Figure~\ref{fig:Hern_agama_comp_ana}, we show the relative error in the potential and forces for both the HO and CylSP methods for the analytical Hernquist models. We note that the CylSP method is more accurate than the HO method for nearly all cases, all shapes, and all radial ranges.

\begin{figure*}
\includegraphics[angle=0, width=180mm]{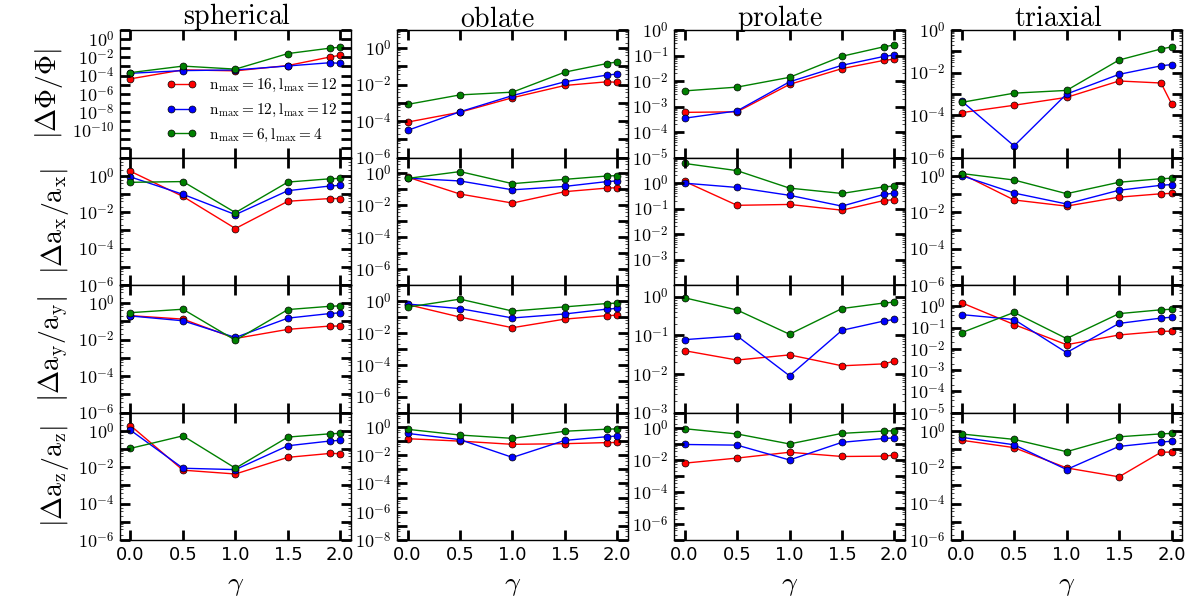}
\caption{Relative errors of the potential and forces for models with a different $\gamma$ value for $r=0.01$. From the left to right, the input density model is spherical, oblate, prolate, and triaxial, respectively. The red, blue, and green lines are the results for the HO method with ($n_{\rm max}$, $l_{\rm max}$)=(16,12), (12,12), and (6,4), respectively.}
\label{fig:Hern_HO_gama}
\end{figure*}

Figure~\ref{fig:Hern_HO_gama} shows the relative errors in the potential and forces for the models with a different inner density profile index $\gamma$, using the HO method for the small radius $r=0.01$. 
It is important to notice that the accuracy of the HO method depends on $\gamma$. In the spherical, oblate, and triaxial systems, the accuracy of $\gamma=1$ is, in general, higher than for the other $\gamma$ values, with the exception of the $z$ component of the force in the triaxial model. This is not surprising since the HO method is designed for $\gamma=1$.     
For models with another value for $\gamma$, the HO method is not accurate for either the potential or the force; this is especially the case for a low number of expansion coefficients. The deviation from the exact values  is $10\%-100\%$ in the force for modes with $\gamma\neq1$. This deviation is due to the HO method itself, and it was also presented in HO92.  For the Hernquist model, the accuracy for the spherical distribution is higher than that for the prolate one. 
We note that there is a significant difference of the accuracy between $\gamma=1.9$ and 2 for the triaxial model with $(n_{\rm max},l_{\rm max})=(16,12)$ in the potential. 
These relative errors depend on the accuracy of the true value and the HO results, and their accuracy 
%From  ~\cite{1996ApJ...460..136M}, we know that the true value arises from a one-dimensional integration and the HO result is from a three-dimensional integration (See HO92).
depends on the density distribution of the model; even two models with similar values for $\gamma$ may exhibit a significant difference. Moreover, we also tested the relative errors in the force for the triaxial models with $\gamma$=1.9, 1.92, 1.94, 1.96, and 2 with $(n_{\rm max},l_{\rm max})=(16,12)$, and we found that these errors between two values for $\gamma$ that are close can differ quite markedly.

\begin{figure*}
\includegraphics[angle=0, width=180mm]{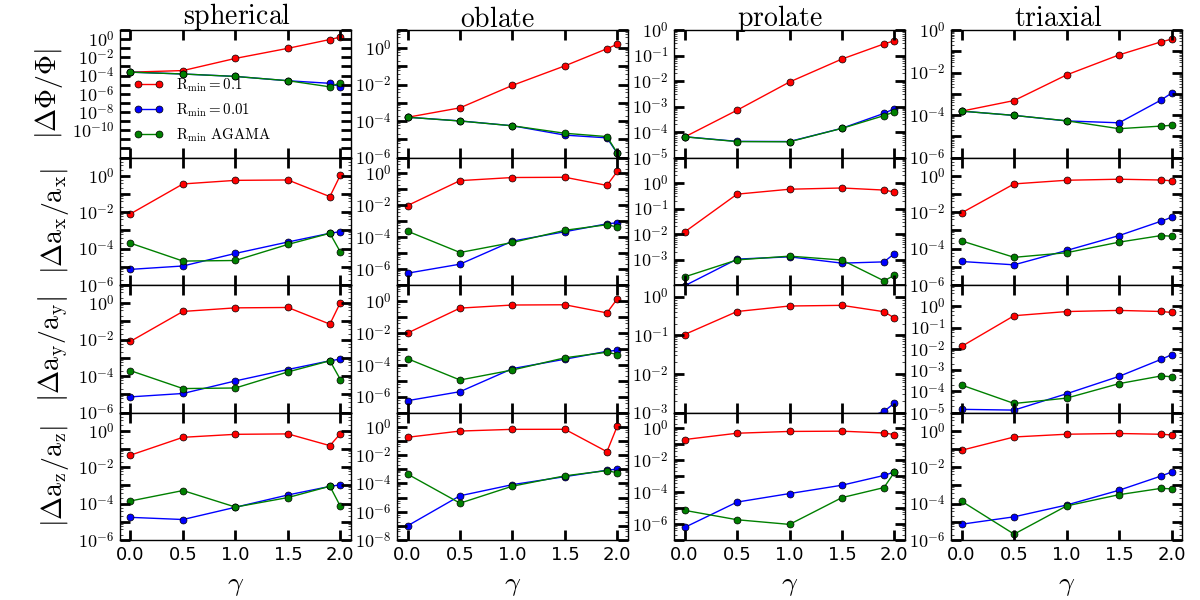}
\caption{Same as in Figure ~\ref{fig:Hern_HO_gama}, but for the CylSP method. The red, blue, and green lines represent the results with $R_{\rm min}=0.1$, 0.01, and the value provided by AGAMA, respectively.}
\label{fig:Hern_agama_gama}
\end{figure*}

Figure~\ref{fig:Hern_agama_gama} shows the relative errors in the potential and forces for the models with a different value for $\gamma$, using the CylSP method for $r=0.01$. 
It is seen that the CylSP method can approximate the potential and forces well with $\gamma=[0,2]$ if $R_{\rm min}=0.01$, or for the small value given by the AGAMA software. This shows that 0.01 and 0 are the appropriate values for $R_{\rm min}$ for the Dehnen models, while $R_{\rm min}=0.1$ is not. 
%It is also noted that the accuracy of forces decreases with $\gamma$ for the spherical, oblate and triaxial models and in $a_y$ and $a_z$ in the oblate models.
Compared with the accuracy from the HO method, the CylSP method performs considerably better at $r=0.01$,  that is, close to the center.
In the spherical and oblate models, we found that there is a significant difference in the accuracy between $\gamma=1.9$ and 2 for $R_{\rm min }$ provided by AGAMA and 0.1, while this difference is small for $R_{\rm min}=0.01$. It is known that forces from the CylSP method are obtained by spline interpolation and that their values are strongly related to the positions of the grids. Moreover, the grids are divided by $R_{\rm min}$, $R_{\rm max}$ and $N_R$, and each grid cell has roughly the same mass~\citep{2013MNRAS.434.3174V}. In other words, the grids depend on the mass distribution of the model. Therefore, even when $R_{\rm min}$, $R_{\rm max}$, and $N_R$ are the same for two different models, the grid positions can be quite different. This is the reason why we find significant differences in the forces for models $\gamma=1.9$ and 2 between $R_{\rm min}=0.1$ and the value provide by AGAMA. Furthermore, we checked the relative errors in the forces for the spherical and oblate models with $\gamma$=1.9, 1.92, 1.94, 1.96, and 2 with $R_{\rm min}=0.1$ and the value given by AGAMA, and we found that the relative errors between two similar $\gamma$ values can also be different. 
Although, both the HO and CylSP methods have sudden jumps from models with $\gamma=1.9$ to 2;  the amplitude of the jump can be much bigger for the HO method than for the CylSP method. For the HO method, it can reach on the order of $100\%$ for $n_{\rm max}=6$, $l_{\rm max}=4$; whereas, in the CylSP method, it can go between $10^{-3}$ and $10^{-4}$, or, in the worst case scenario, between $10^{-3}$ and $10^{-2}$. Thus, here as well the CylSP method performs much better than the HO method.

%It is seen that the CylSP method can approximate the potential for model with $\gamma=[0,2]$ and force well for $\gamma\leq1.5$ at $r=0.01$. For models with large $\gamma>1.5$, the CylSP method can give some errors for force.

\begin{figure}
\includegraphics[angle=0, width=90mm]{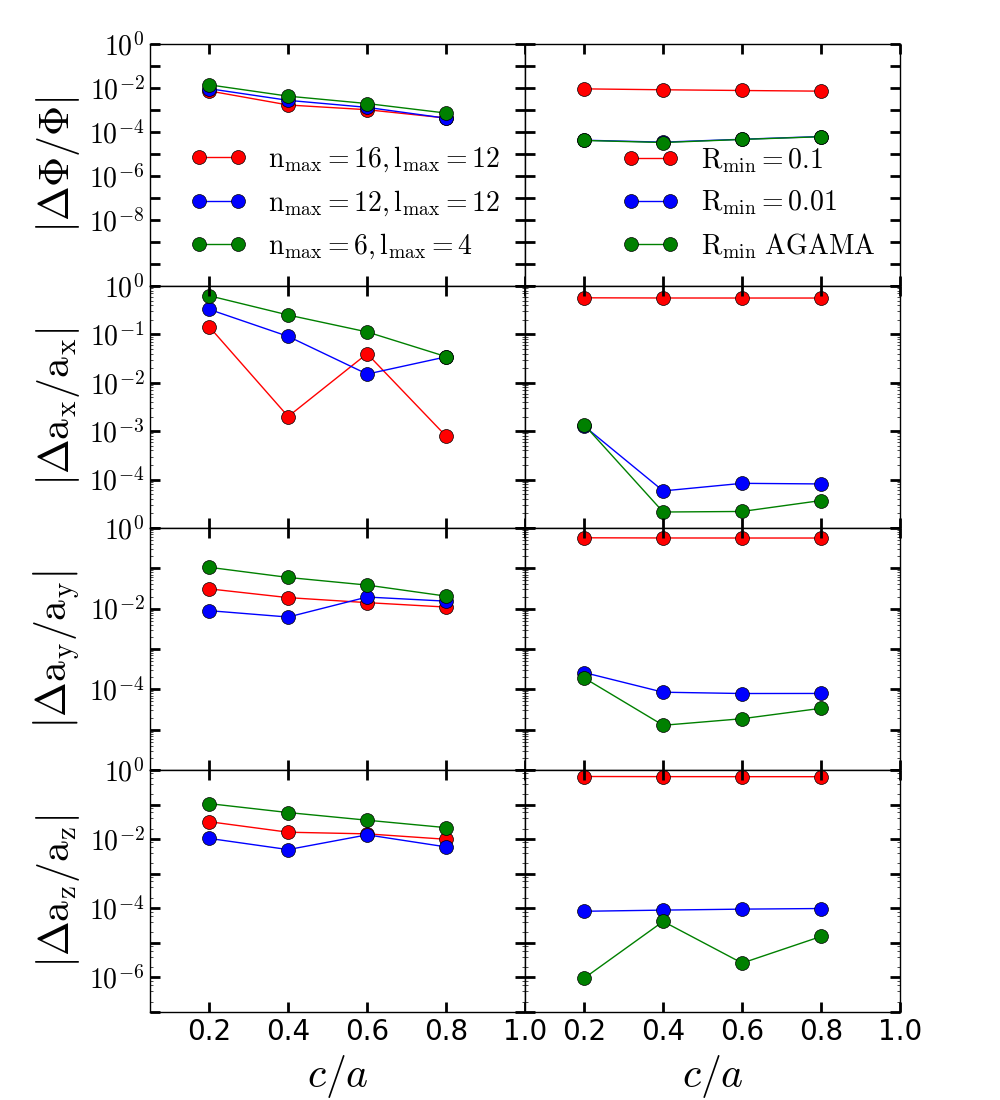}
\caption{Relative error in potential and forces for both the HO ($left$) and the CylSP methods ( $right$) for the prolate  models with a different $c/a$ at $r=0.01$.
In the left panels, the red, blue, and green lines are the results for the HO method with ($n_{\rm max}$, $l_{\rm max}$)=(16,12), (12,12), and (6,4), respectively.
In the right panels, the red, blue, and green lines represent the results for the CylSP method with $R_{\rm min}=0.1$, 0.01, and the value given by AGAMA, respectively. From top to bottom, the radius $r$ is along the $x-$, $x-$, $y-$, and $z-$ axes, respectively.}
\label{fig:Hern_HO_ca}
\end{figure}

The left panels of Figure~\ref{fig:Hern_HO_ca} show the relative error of the potential and forces for prolate models with a different $c/a$ by using the HO method for the small radius $r=0.01$. We find that the HO method can give a better estimation for the potential than for the forces. For the latter, the HO method gives more accurate values for larger values of $c/a$. The reason is that the model becomes more spherical with an increasing $c/a$ and that a spherical system has a higher accuracy than a prolate one, as is shown in Figure~\ref{fig:Hern_HO_gama}.

The right panels of Figure~\ref{fig:Hern_HO_ca} show the relative error of the potential and forces for prolate models with a different $c/a$ by using the CylSP method for the small radius $r=0.01$. It is found that the CylSP method can give accurate values for both the potential and forces in the prolate model at $r=0.01$ and {\bf the accuracy is nearly the same  with $c/a\geqslant 0.4$ in the potential, $a_x$ and $a_y$.} We can thus conclude that the CylSP method is more accurate than the HO method at $r=0.01$. 

\subsection{Comparisons for particle realizations of generalized Dehnen potentials}

\begin{figure*}
\includegraphics[angle=0, width=180mm]{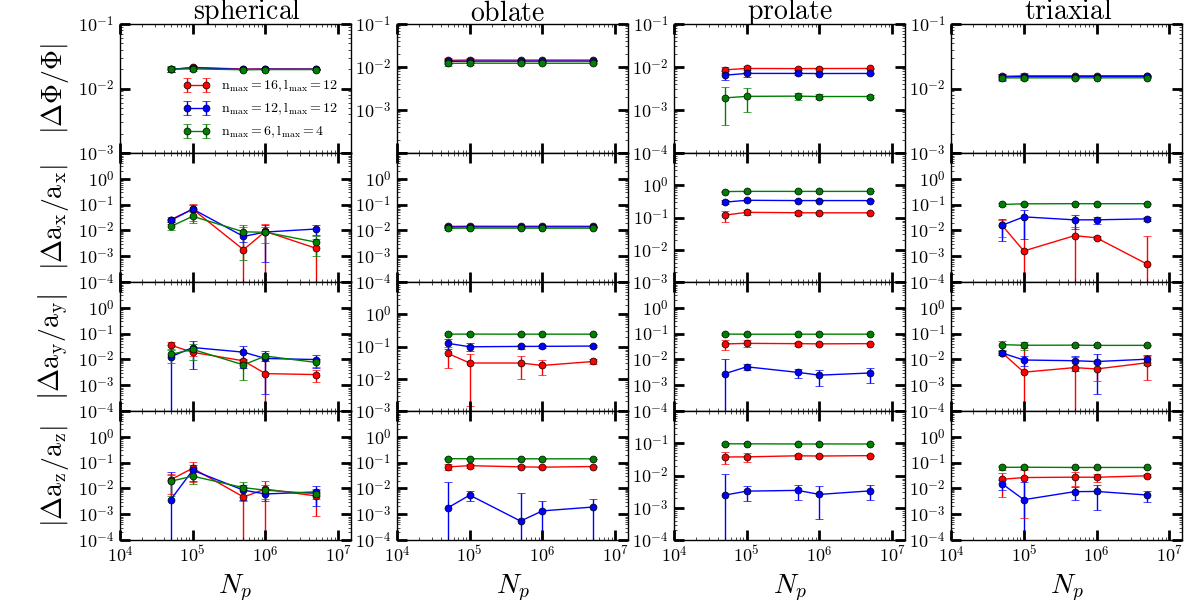}
\caption{ Relative errors of the potential and forces for the Hernquist models  with different particle numbers at $r=0.01$ when using the HO method. In each panel, the red, blue, and green lines are the results for HO method with ($n_{\rm max}$, $l_{\rm max}$)=(16,12), (12,12), and (6,4), respectively. The filled circles are the mean value from 40 random samples, and the error bar is the standard error from these samples.}
\label{fig:Hern_HO_np}
\end{figure*}

In order to study the accuracy of the two expansion methods for particle realizations of given potentials, we used the inverse function sampling method to generate the Dehnen models with spherical, oblate, prolate, and triaxial shapes as well as different particle numbers. Figure~\ref{fig:Hern_HO_np} shows the relative errors of the potential and forces for the Hernquist models with different particle numbers using the HO method at radius $r=0.01$. We randomly generated 40 samples with different particle numbers. The filled circles in Figure~\ref{fig:Hern_HO_np}  are the mean value and the error bars are the standard error from these 40 samples. We also note that the error bar size in each point is different.
One possible reason is that  there is a high Possion noise level if the particle number is smaller than $10^6$, particularly in the central region.  In the central region $r\le0.01$, the density is high but the total enclosed mass is low, resulting in the Poisson noise due to the small number of particles sampled. The second reason is that the error bar size also depends on the mean value of the deviation since we use a logarithmic scale for the $x$-axis.   

It is seen that the forces obtained from the HO method deviate from the true value less than $2\%$ if $N_p\ge 5\times10^5$, and this deviation is nearly independent of the number of expansion coefficients $n_{\rm max}$ and $l_{\rm max}$.   
For the oblate model,  the force deviations for the HO method with ($n_{\rm max}$, $l_{\rm max}$)=(16,12) are smaller than $\sim 9\%$ and these deviations weakly depend on the particle number. It is also noted that the forces along the $z$-axis with 
 ($n_{\rm max}$, $l_{\rm max}$)=(12,12) are more accurate than those with ($n_{\rm max}$, $l_{\rm max}$)=(16,12). As shown in HO92, increasing the number of expansion terms can improve the accuracy of the system as a whole, the reconstructed acceleration with a higher number of expansion coefficients can deviate more from the corresponding exact values at the central region (HO92). This is one possible reason as to why we find the results from a lower number of expansion coefficients at r = 0.01 that are more accurate than those from a higher number of expansion coefficients.  Another possible reason is that there is a high Poisson noise level in the central region.      
The prolate model has the largest force deviation of 
these four shapes. These deviations are $\sim 20\%$ for ($n_{\rm max}$, $l_{\rm max}$)=(16,12) and these values are nearly constant along the particle numbers. 
The triaxial model is more accurate than the oblate and prolate models, but less accurate than the spherical one.
The force deviations are $\sim3\%$ for ($n_{\rm max}$, $l_{\rm max}$)=(12,12)  at $r=0.01$ and they depend only weakly on the particle number.

\begin{figure*}
\includegraphics[angle=0, width=180mm]{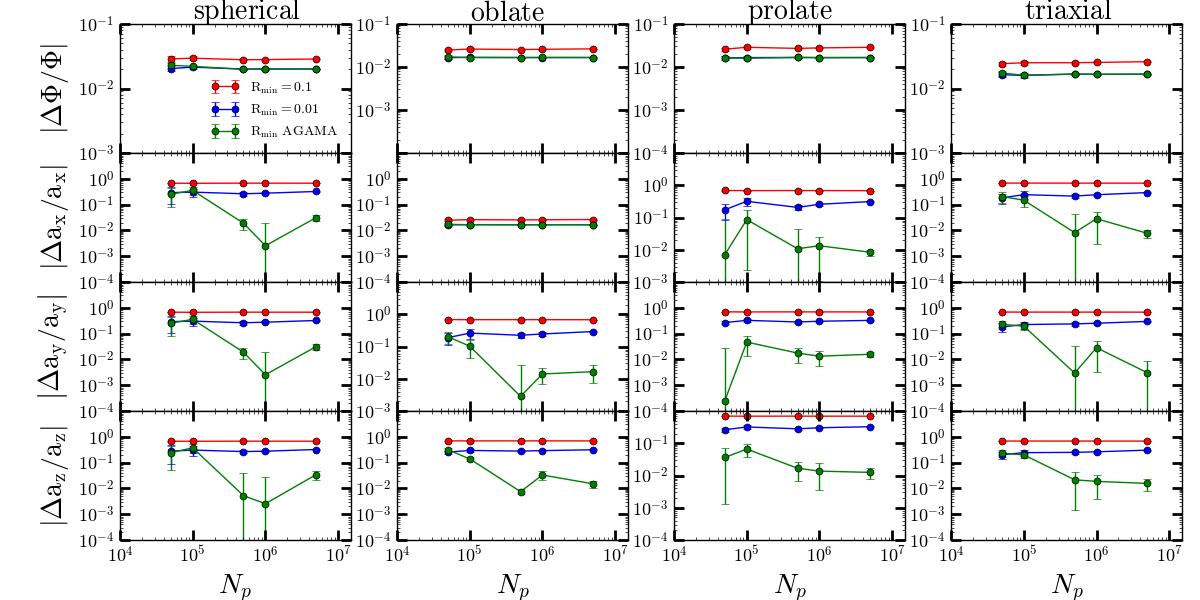}
\caption{Same as in Figure~\ref{fig:Hern_HO_np}, but for the CylSP method. The red, blue, and green lines represent the results in the CylSP method for $R_{\rm min}=0.1$, 0.01, and the value given by AGAMA, respectively.} 
\label{fig:Hern_agama_np}
\end{figure*}

Figure~\ref{fig:Hern_agama_np} shows the relative errors of the potential and forces for the Hernquist models with different particle numbers using the CylSP method at radius $r=0.01$.  It is clear that the force accuracy for the $R_{\rm min}$ provided by the AGAMA software is higher than that for $R_{\rm min}$=0.01 and 0.1; additionally, this accuracy increases with the particle number. This is because the CylSP method uses the spline interpolation method, and the accuracy from the interpolation is higher than that from the extrapolation. A larger number of particles can reduce the Poisson noise of the system and reproduce the density distribution of the system better. 

We see that the force accuracy for the CylSP method with $R_{\min}$ as given by the AGAMA software is equivalent to that of the HO method for the spherical and triaxial models, and higher than the HO method in the oblate and prolate models with $N_p\ge 5\times10^5$.  
Compared to the analytical model, we can see that the advantages of the CylSP method over the HO method are weakened for the particle samples, as was also found by ~\cite{2015MNRAS.450.2842V}.  The reason is that, in the analytical models,
the HO method expansion coefficients are determined by a three-dimensional integration while the expansion coefficients in the CylSP method are obtained from a two-dimensional integration. 
For particle realizations, however, the integration is unnecessary for both methods in calculating the expansion coefficients. 

We also compared the accuracy for these two methods at a smaller inner radius, namely $r=0.001$. We found that the accuracy for forces for the HO method is higher than that for the CylSP method in the spherical and triaxial models and this trend is independent of the particle number.  For the oblate and prolate models, the CylSP method is more accurate than the HO method if $N_p\geqslant10^6$.        

Except for the force accuracy, the required CPU time is also important for the orbit integration. We consider both the initialization time, which was only performed only once, and the subsequent time. In the HO method, the initialization part calculates the expansion coefficients, while both the expansion coefficients and the force on the grid are calculated in the CylSP method. For an analytical model, the expansion coefficients are determined by three-dimensional integrations in the HO method, while those are calculated by two-dimensional integrations in the CylSP method. If the CPU time for the force on the fixed grid is also considered, the initialization time for the CylSP method can  be comparable to that in the HO method for the analytical model. In this paper, we study both the initialization time for an $N$-body realization and the time after initialization to calculate the force and acceleration.   

\begin{figure}
\includegraphics[angle=0, width=92mm]{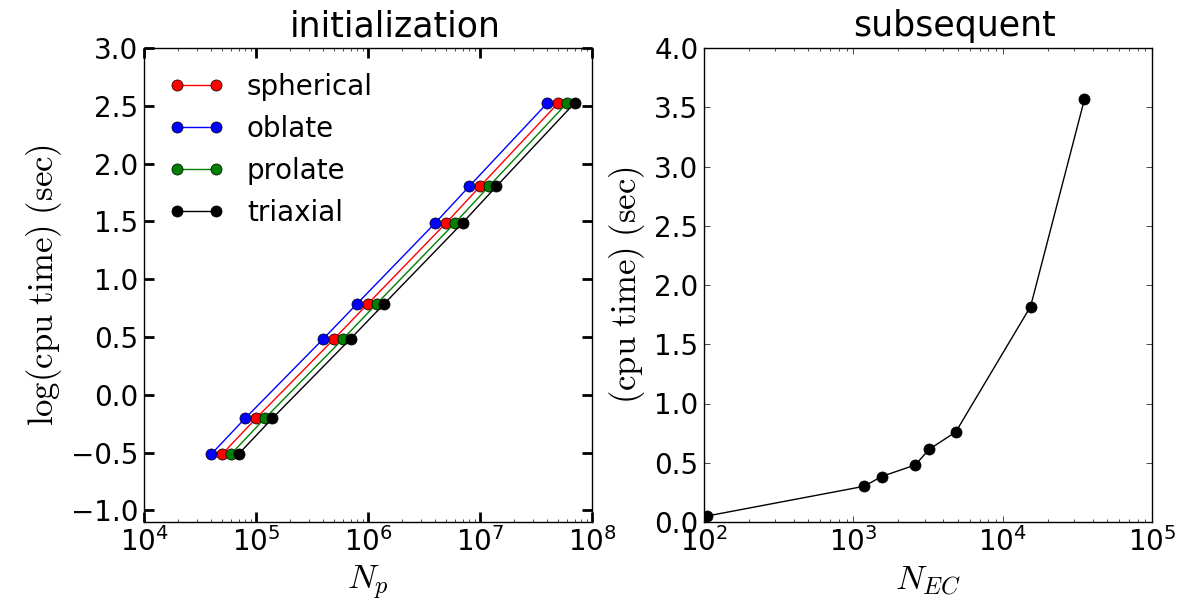}
\caption{CPU time for the HO method. Left: Dependence of the initialization CPU time on the particle number of a realization of the Hernquist model. The red, blue, green, and black lines are the results for the spherical, oblate, prolate, and triaxial models, respectively. The blue, green, and black lines are slightly shifted along the $x$-axis.
Right: Dependence of the subsequent CPU time on the number of the expansion coefficients. The subsequent CPU time is estimated by the forces at 10,000 positions.}
\label{fig:cpu_HO}
\end{figure}

On the left side of Figure~\ref{fig:cpu_HO}, we show the dependence of  the initialization CPU time on the particle number from the HO method. It is clear that the initialization time is proportional to the particle number and independent of the model shape. 
This is normal because the expansion coefficients in the HO method are the sum of a function of the particle position (see Eq. 3.17 in ~\citealt{1992ApJ...386..375H}). 
In the HO method, the subsequent time is used to sum the expansion terms, which are related to the number of the expansion coefficients $N_{EC}=\frac{1}{2}(n_{\rm max}+1)(l_{\rm max}+2)(l_{\rm max}+1)$. For both the analytical model and $N$-body particles, once the expansion coefficients have been obtained, the subsequent CPU time is the same if the same $N_{EC}$ is adopted. The right panel of  Figure~\ref{fig:cpu_HO} shows the dependence of the subsequent CPU time on the number of the expansion coefficients for the HO method.  We find that the subsequent time in the HO method increases significantly with the number of the expansion coefficients, which can be explained by Eq. (3.17) in HO92.

\begin{figure}
\includegraphics[angle=0, width=90mm]{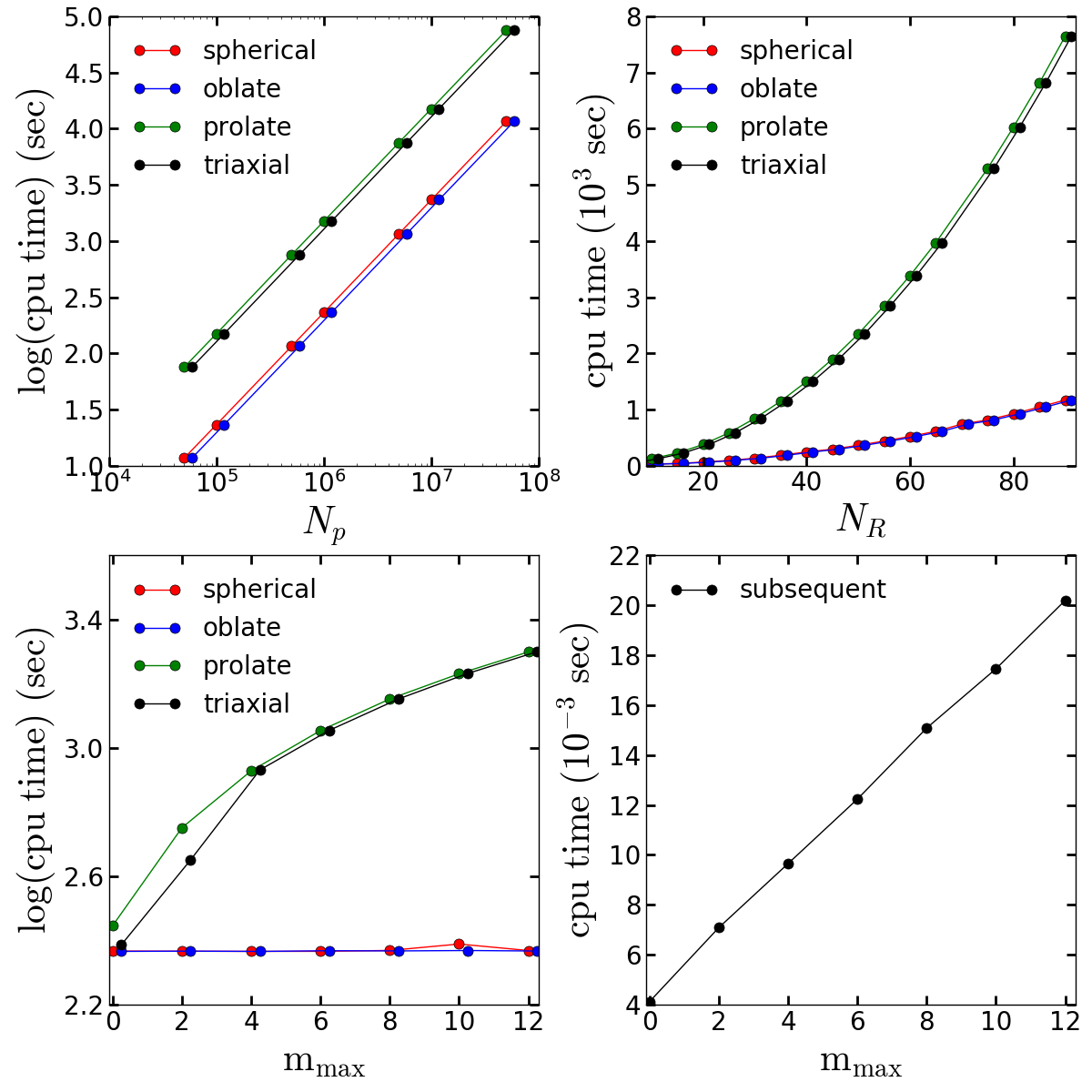}
\caption{CPU time for the CylSP method. Top left: Dependence of  the initialization CPU time on the particle number of the Hernquist model realization. Top right: Dependence of the initialization CPU time on the parameter $N_R$ for an $N$-body realization of the Hernquist model with $N_p=10^6$.  Bottom left: Dependence of the initialization CPU time on the parameter $m_{\rm max}$ for a snapshot of an $N$-body realization of the Hernquist model with $N_p=10^6$.   Bottom right: Dependence of the subsequent CPU time on the parameter $m_{\rm max}$.  The subsequent CPU time is estimated by the forces at 10,000 positions.
Excluding the bottom right panel, the blue and black lines are slightly shifted to the right along the $x$-axis and the red, blue, green, and black lines are the results of the spherical, oblate, prolate, and triaxial models, respectively.} 
\label{fig:cpu_agama}
\end{figure}

Figure~\ref{fig:cpu_agama} shows the dependence of the CPU time on the parameters of particle number $N_p$ (top left), and $N_R$ (top right) and $m_{\rm max}$ (two bottom panels) for the CylSP method. The bottom panel is for the subsequent calculation time while the other three panels are for the initialization CPU time.  We find that the initialization CPU time increases significantly with the particle number $N_P$ and the grid number $N_R$.  It is noted that the initialization CPU time in the oblate and spherical models is significantly smaller than that in the prolate and triaxial models. The reason is that only the $m=0$ terms are needed for a disk system,  with the spherical system being a special ``disk" system. This is the reason why we find the initialization CPU time to be a constant in the spherical and disk models, which can be seen in the bottom left panel of Figure~\ref{fig:cpu_agama}. 
 
Compared with the HO method, the initialization CPU time in the CylSP is high. However, the subsequent time in the CylSP method is much lower than that in the HO method. The reason is that the subsequent time in the HO method is the cost of summing $N_{CE}$ terms of the basis function; whereas for the CylSP method, it is only the time to use a two-dimensional interpolation routine ~\citep{2019MNRAS.482.1525V} .Therefore, if a large number of orbits need to be calculated, such as millions of orbits, this is something that should be kept in mind.      

%However, if the input model is an $N$-body snapshot the CPU time from the CylSP method is significantly larger than that from the HO method. For example, the CPU time for the CylSP is 45 times of that for the HO method for a spherical model with $10^6$ %particles, and this difference increases with particle number. Since the spline interpolation method is adopted in CylSP, once the potential and force in the fixed grids are calculated, the force and potential in any position can be obtained from the values of these fixed %grids, which saves the CPU time in the subsequent computations  in the force and potential during the orbit integration. In other words, the advantage of the speed from the HO method will gradually be lost with increasing orbit number. }      

\section{A multicomponent model with a Ferrers bar}
Since nearly two-thirds of spiral galaxies in the Universe are barred \citep[e.g.,][]{2012ApJ...745..125L, 2010ApJS..190..147B, 2015ApJS..217...32B}, it is important to discuss the application of the different expansion methods to barred galaxies.
Here we consider a fiducial barred galaxy model, which consists of a Miyamoto (MN) disk ~\citep{1975PASJ...27..533M},  a Plummer halo ~\citep{1911MNRAS..71..460P} and a Ferrers bar ~\citep{1877QJPAM..14....1F}. The parameters we used are those listed in Table 1 in ~\cite{2002MNRAS.333..847S} and, for reasons of continuity, we have the same bar orientation, that is, along the $y$ axis. The potential and force for this bar model can be obtained directly by following the Appendix in ~\cite{1984A&A...134..373P}.

\begin{figure*}
\includegraphics[angle=0, width=180mm]{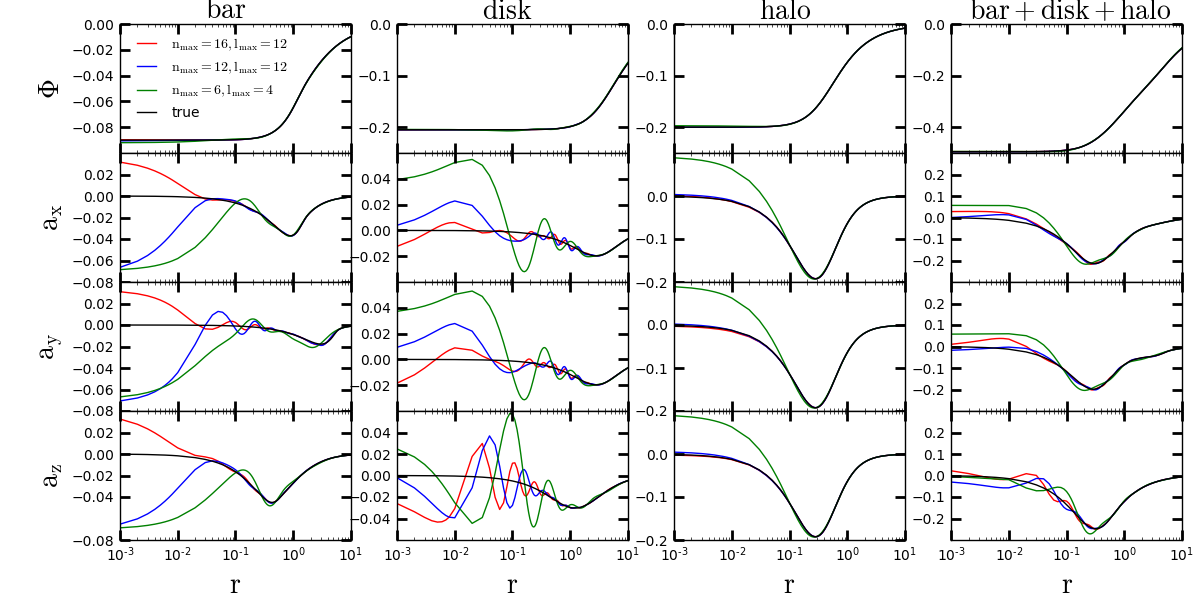}
\caption{Potential and forces for the barred galaxy models and their individual components. From left to right, the input density model is the Ferrers  bar, the MN disk, the Plummer halo, and the total, i.e., bar+disk+halo, respectively. In each panel, the black line
corresponds to the results for the model given by the analytical method directly. The red, blue, and green lines are the results for the HO method with ($n_{\rm max}$, $l_{\rm max}$)=(16,12), (12,12), and (6,4), respectively. }
\label{fig:bar_HO_ana}
\end{figure*}

Figure~\ref{fig:bar_HO_ana} shows the potential and forces for the bar models using the HO method. The input model is an analytical model.  It is seen that the HO method can reconstruct the potential well, except for the potential in the central region ($r<0.1$) of the Ferrers bar and the MN disk.  For the Plummer halo, the HO method can give a good approximation if a large number of expansion coefficients are used.  

\begin{figure*}
\includegraphics[angle=0, width=180mm]{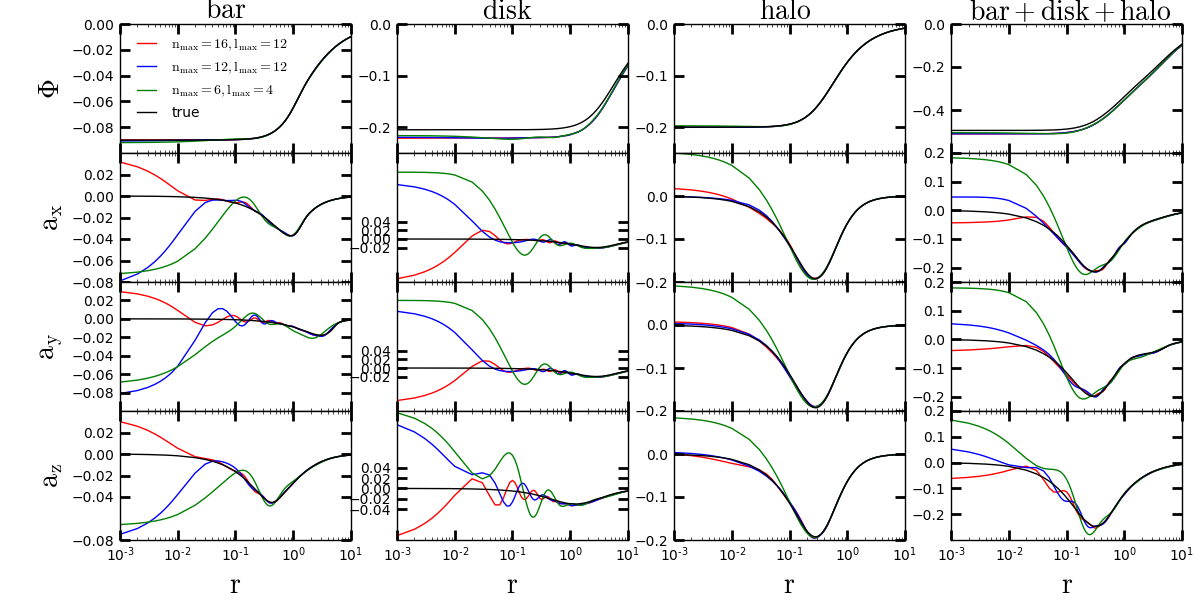}
\caption{Same as in Figure~\ref{fig:bar_HO_ana}, but the only difference is that the input is an $N$-body realization of our barred galaxy model. The total particle number for the bar is $2\times10^6$. The particle numbers for the bar, disk, and halo are $200000$, $1640000$, and $160000$, respectively. }
\label{fig:bar_HO_sim}
\end{figure*}

Figure~\ref{fig:bar_HO_sim} also shows the potential and forces for the models using the HO method, but now the input is an $N$-body realization of the model. It shows again that the HO method can provide large errors for the force in the central region, both for the Ferrers bar and the MN disk. It is also noted that the potential of the MN disk, as calculated by the HO method,  significantly deviates from the true value. One possible reason was that our input $N$-body snapshot is sampled using the revised function method.  We have compared the density from the sampled particles with the analytical model; the density at $r<1$ has a $5\%$ deviation for the sampled particles, as was also found by ~\cite{2015MNRAS.450.2842V}.

\begin{figure*}
\includegraphics[angle=0, width=180mm]{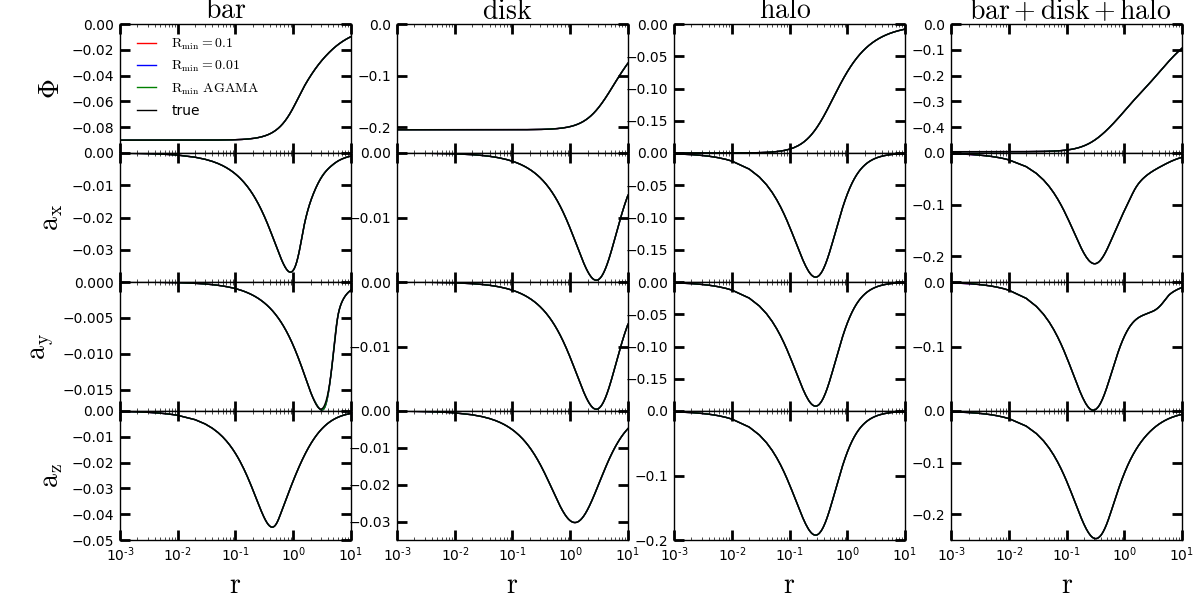}
\caption{Same as in Figure~\ref{fig:bar_HO_ana}, but for the CylSP method. The red, blue, and green lines represent the result in the CylSP method with $R_{\rm min}=0.1$, 0.01, and the value provided by AGAMA, respectively. 
Since the results from three $R_{\rm min}$ are the same as the true values, only the black lines can been seen in most panels.}
\label{fig:bar_agama_ana}
\end{figure*}

Figure~\ref{fig:bar_agama_ana} shows the potential and forces for the analytical barred galaxy models using the CylSP method. Clearly, CylSP can reproduce the potential and forces well for the Ferrers  bar, MN disk, and the Plummer halo.
%There is one exception, however, namely for the force along the $y-$axis. We note that the form of the curve does not change, but the location of the maximum is about $15\%$ shorter and the maximum itself is $11\%$ lower, consistent with what was found by ~\cite{2015MNRAS.450.2842V} We can, therefore, argue that the bar found by CylSP is simply a somewhat different bar, which is somewhat shorter and stronger than the true bar. 
%This will not introduce qualitative differences in the orbital structure, nor even major quantitative ones, contrary to the errors of the HO method which introduce a discontinuity in the center, that can have considerable effects on the orbital structure, e.g. by introducing chaos.
We also find that the results from different $R_{\rm min}$ are nearly same;  the reason is that the potential and accelerations vary slowly in the central region in these models.   

\begin{figure*}
\includegraphics[angle=0, width=180mm]{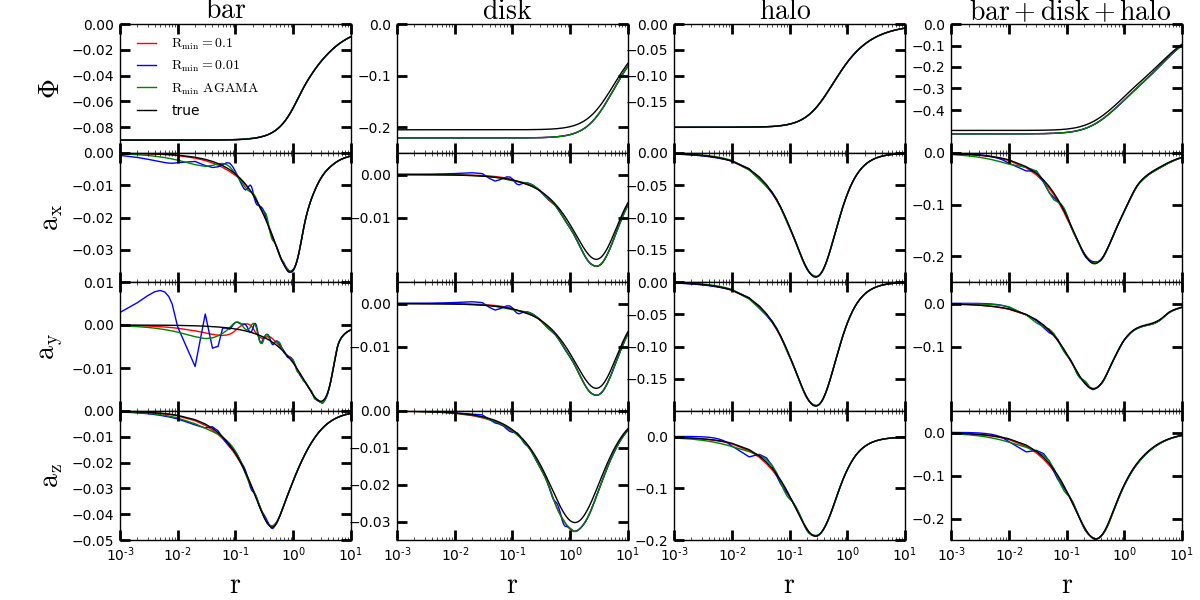}
\caption{Same as in Figure~\ref{fig:bar_agama_ana}, but the only difference is that the input is an $N$-body realization of our barred galaxy model. The total number of particles is $2\times10^6$. }
\label{fig:bar_agama_sim}
\end{figure*}

\begin{figure*}
\includegraphics[angle=0, width=180mm]{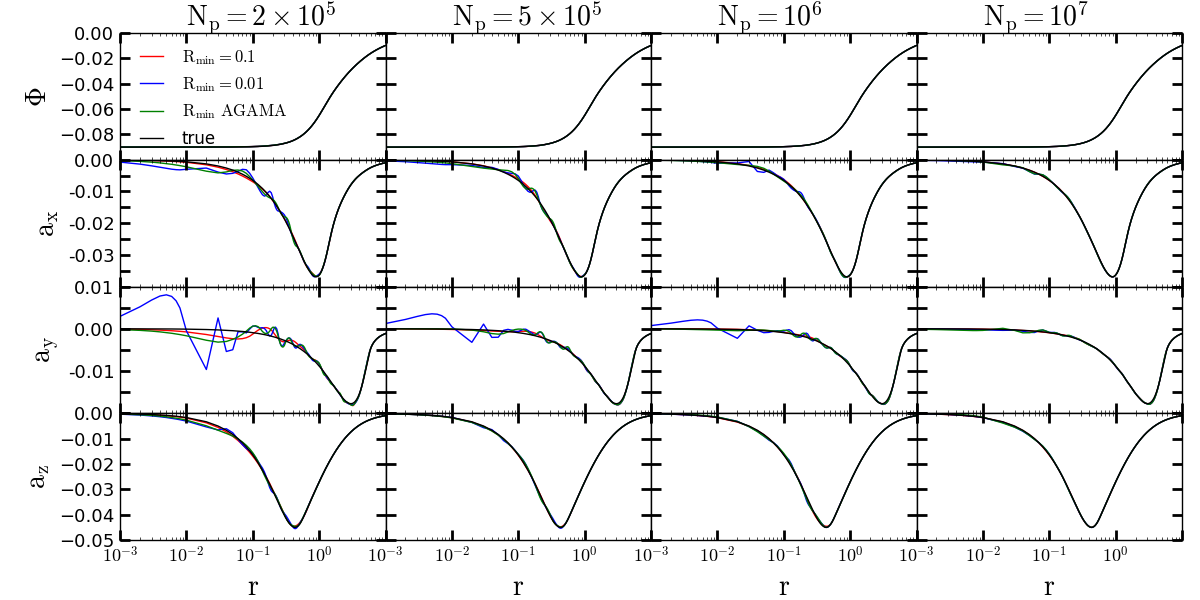}
\caption{Potential and forces for the Ferrers bar with different  redistribution numbers of particles. In each panel, the black line is the results for the model given by the analytical method. The red, blue and green lines represent the result in the CylSP method with $R_{\rm min}=0.1$, $R_{\rm min}=0.01$ and the value provided by AGAMA, respectively. From left to right, the results for $N$-body realizations with particle number $2\times 10^5$, $5\times 10^5$, $10^6$, and $10^7$, respectively.}
\label{fig:ferrer_agama_sim}
\end{figure*}

Figure~\ref{fig:bar_agama_sim} shows the potential and forces for the barred galaxy models by using the CylSP method. The input is the particle  realization with particle number $N_p=10^6$.
Generally, we find that the CylSP method is more accurate than the HO method, especially in the central region of the system. However, there are some wiggles in the force distribution in the Ferrers  bar, MN disk, and Plummer halo, and the results from $R_{\rm min}=0.01$ and the $R_{\rm min}$ provided by AGAMA give larger wiggles than those from $R_{\rm min}=0.1$.  These wiggles affect the orbit shape in the system, especially the box orbits. To determine the reason why these wiggles occur, we increased the number of particles
in the $N$-body model. As is shown in Figure~\ref{fig:ferrer_agama_sim}, the wiggles become weaker and weaker with the increasing particles and the wiggles vanish if the particle number is $10^7$.

 %\begin{figure}
%\includegraphics[angle=0, width=80mm]{bar_HO_agama_comp_ana.png}
%\caption{ Relative error in potential and forces for both the HO (red line) and CylSP (blue line) methods for the  bar model. For the HO method, $[n_{\rm max}=16$, $l_{\rm max}=12]$. For the CylSP method, $m_{\rm max}=4$, $N_R=25$, $N_z$=25, $r_{\rm min}=0.1$.}
%\label{fig:HO_agama_comp_ana}
%\end{figure}

 %\begin{figure}
%\includegraphics[angle=0, width=80mm]{bar_HO_agama_comp_sim.png}
%\caption{Similar to Figure~\ref{fig:HO_agama_comp_ana}, the only difference is that the input is an $N$-body model with $10^6$ particles.}
%\label{fig:HO_agama_comp_sim}
%\end{figure}

\begin{figure*}
\includegraphics[angle=0, width=180mm]{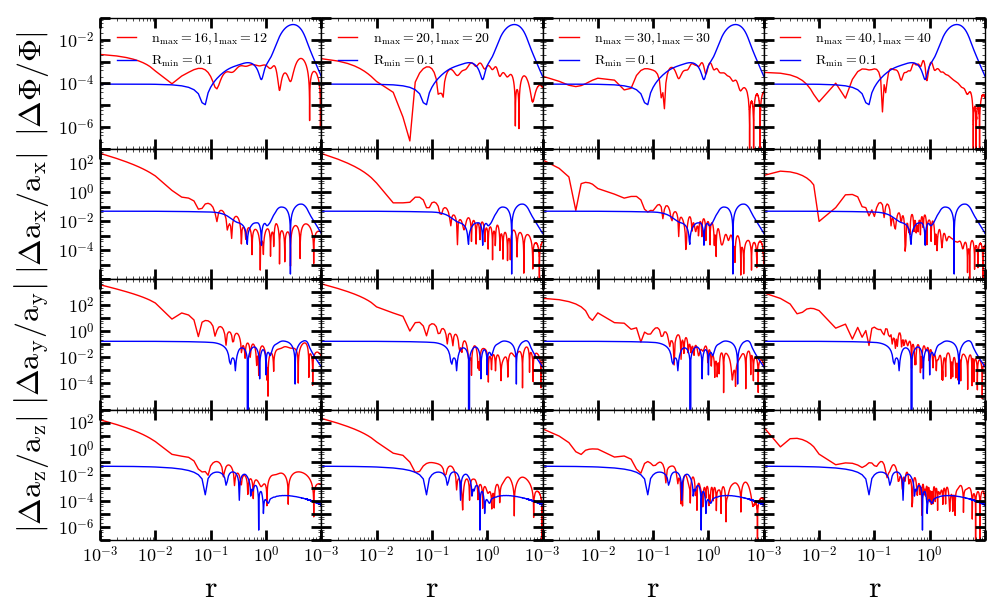}
\caption{Relative errors of potential and forces for the HO (red lines) and CylSP (blue lines) methods for snapshots of an $N$-body realization of a Ferrers bar with a particle number $N_p=10^6$. From left to right, the results for the HO with 
($n_{\rm max}$, $l_{\rm max}$)=(16,12), (20,20), (30,30) and (40,40), respectively. 
For the CylSP method, $R_{\rm min}=0.1$, $N_R=40$, $N_z=40$, $z_{\rm min}=0.1$, $R_{\rm max}=z_{\rm max}=100$.}
\label{fig:comp_nmax}
\end{figure*}

For the Ferrers bar model, we have found that the CylSP method is more accurate than the HO method along the radius even when the expansion coefficient terms with $n_{\rm max}=16$, $l_{\rm max}=12$ are adopted. As is shown in Figure~\ref{fig:comp_nmax}, the accuracy of the HO method can be comparable to the CylSP method at large radii ($r>0.1$) if $n_{\rm max}=40$, $l_{\rm max}=40$ are used. However, the accuracy in the central region is not improved for the HO method. 
%The cost in CPU time in the HO method with parameters $n_{\rm max}=40$, $l_{\rm max}=40$ is one-thirteenth of the CylSP method.   

Since the accuracy of the forces from the HO and CylSP methods is different, this will affect the accuracy of the orbit integration, and thus the orbit structure. One useful way to characterize  the orbit structure is the surface of section (SOS).  In the bar model, we follow ~\cite{2002MNRAS.333..847S} and consider the $y$-axis to be the longest major axis. Therefore, we consider the intersection of the orbit with the plane y=0 and in particular those with $v_y>0$. For each energy, we generated 25 initial conditions of  orbits with $x\neq 0$ and $v_y\neq 0$.

\begin{figure*}
\includegraphics[angle=0, width=180mm]{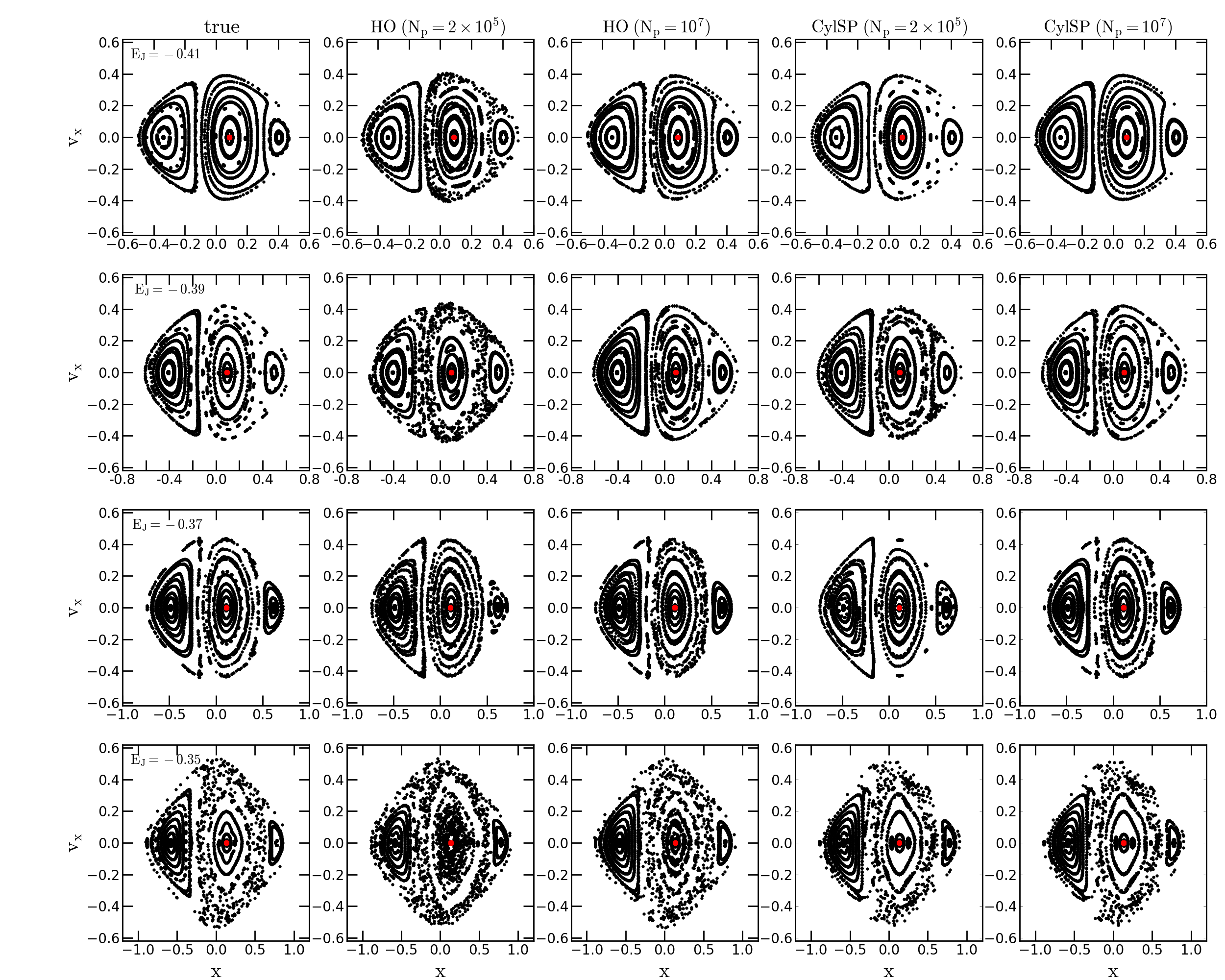}
\caption{Surfaces of section (SOS) for the bar model. The left column gives the results for the analytic potential. The second and third columns show the results for the HO method with $N_p=2\times10^5$ and $N_p=10^7$, respectively. The fourth and fifth columns show the results for the CylSP method with $N_p=2\times10^5$ and $N_p=10^7$, respectively. 
From top to bottom, the rows correspond to $E_J=-0.41$, -0.39, -0.37, and -0.35, respectively. The red filled circle in each panel is the position of the x1 orbit.}
\label{fig:sos}
\end{figure*}

Figure~\ref{fig:sos} shows the SOS for four Jacobi's energies from the analytical potential, the HO method and the CylSP method, respectively. In order to focus on studying how the accuracy of the force from the bar affects the SOS, we use
analytical force and acceleration calculations for the MN disk and the Plummer halo. We also consider two different particle numbers of the bar to probe the discreteness noise.           
It is seen that the overall framework of the SOS at each $E_J$ from three methods are similar, but the differences in the details are also obvious. 

For example, around the x1 orbit, more chaotic orbits are generated from the HO method than from the true and CylSP method.      
This can be due to higher numerical errors in the central region, which we already discussed above. This can introduce a discontinuity and therefore chaos. It is also clear that the discreteness noise can be reduced with increasing particle number in both methods. However, even when the particle number is $10^7$, the HO method still has large errors in the central region, and thus more chaotic orbits appear in the central region than those in the true and CylSP method. For the CylSP method, the $N$-body bar model with $N_p=10^7$ can reconstruct almost the same SOS as the true potential, which again shows that the CylSP method works better than the HO method for a bar. 
 In Table~\ref{table:ej}, we show that the stability of the x1 orbit at four different $E_J$. We found that the stability of the x1 orbit at these $E_J$ are the same for the three methods.
 Since we fixed the disk and halo potential, and as the particle mass of the Ferrers bar only contributes $10\%$ of the mass for the whole system, the effect of stability due to the force accuracy for the three methods is not obvious. 
We also tested the stability of the x1 orbit at $E_J=-0.37$ and $E_J=-0.35$, which could show considerable differences between the three methods if the disk and halo were also represented by particles. 

%Indeed the  CylSP approximation does not change the form of the curve, it just represents a somewhat different bar.  As the position of the maximum increases with the bar length and its maximum increases with the bar strength, we can can say that the bar of the %CylSP approximation is somewhat shorter, of the order of $15\%$, and somewhat stronger, bar of the order of  $11\%$. These numbers are  actually the relative error for  the location of the maximum and the height of the maximum. On the contrary, the HO error %introduces a discontinuity in the center. This is expected to have an effect on the orbital structure and, indeed, as we discussed in the previous section the HO based potential introduces  considerably more chaos than the CylSP one (See Figure~\ref{fig:sos}).

\begin{table*}
\caption{Four Jacobi energies $E_J$ in the bar potential. Row (1): The name of three methods. Row (2): The position of the Lagrangian points $L_{1,2}$.  Row (3): The Jacobi energy at $L_{1,2}$.  Rows: (4)-(7) The stability of the x1 orbit at these Jacobi's energies. Rows: (8)-(11): The ratio of $E_J/E_J,L_{1,2}$.}\label{table:ej}
\begin{tabular}{cccccc}\hline
                   &true & HO ($N_p=2\times10^5$) &HO ($N_p=10^7$) & CylSP ($N_p=2\times10^5$)& CylSP ($N_p=10^7$) \\
 \hline                   
 
  $L_{1,2}$ &  $\pm 6.044056$ &  $\pm 6.045628$ &  $\pm 6.045613 $ & $\pm  6.045552$ &$\pm  6.045552$ \\
 $E_{J,L_{1,2}}$ &-0.195471 &-0.195470  & -0.195473  & -0.195477&-0.195475 \\
 $E_{J,1}=-0.41$& stable & stable & stable &stable &stable \\  
 $E_{J,2}=-0.39$ & stable &stable&stable &stable &stable  \\
 $ E_{J,3}=-0.37$& stable &stable&stable &stable &stable \\
 $ E_{J,4}=-0.35$& unstable &unstable &unstable&unstable &unstable \\  
  $E_{J,1}/E_{J,L_{1,2}}$ & 2.097498  & 2.097509 &2.097476 &2.097433&2.097455\\
  $E_{J,2}/E_{J,L_{1,2}}$ & 1.995809  &1.995191  &1.995160 &1.995120&1.995140  \\
  $E_{J,3}/E_{J,L_{1,2}}$ & 1.892864  &1.892874  &1.892844 &1.892806 &1.892825 \\
  $E_{J,4}/E_{J,L_{1,2}}$ &1.7905469 &1.790556  &1.790529  &1.790492&1.790510 \\

\hline
\end{tabular}
\end{table*}

\section{Summary and discussions} 
The accuracy of the potential and force calculations directly affects the orbit properties in a model. In this paper, we present a detailed comparison of the Hernquist \& Ostriker (1992, HO) method in the spherical coordinate system with the CylSP method~\citep{2015MNRAS.450.2842V},  which uses the  two-dimensional spline interpolation for Fourier coefficients in the cylindrical coordinate system . We have compared these two methods for two kinds of density distribution; one has a Dehnen profile~\citep{1993MNRAS.265..250D}, which is generalized to nonspherical shapes (See Equation~\ref{eq:dehnen}).  If the inner density power-law index $\gamma$ is equal to 1, this is the Hernquist model.  The second kind of model that we consider is a
multicomponent galaxy model, which consists of a Miyamoto disk, a Plummer halo, and a Ferrers bar.

For the Hernquist model (with $\gamma=1$), we find that the CyLSP method is more accurate than the HO method. This is especially the case in the central region if the input model is an analytical density distribution and a reasonable parameter set is chosen in the CyLSP method. However, if the input density is an $N$-body realization, we find that the force accuracy in the CylSP method is more sensitive to the particle number than the HO method. The force accuracy from the CylSP method at the central region is equivalent to the HO method for the spherical and triaxial models, and it is only higher than the HO method in the oblate and prolate models if the particle number is $N_p\ge5\times 10^5$.  If the particle number is smaller than $5\times 10^5$, the HO method is better than the CylSP method for the spherical and triaxial models.  

For the generalized Dehnen model with $\gamma\ne1$,  the force accuracy from the HO method is lower than that for the Hernquist model.  The CylSP method is more accurate than the HO method if the input density is the analytical Dehnen model. For the snapshot of an $N$-body realization, we also checked the force accuracy for these two methods at the central region. The advantage of the CylSP method to the HO method is significant for the oblate shapes. For the more prolate Dehnen model, the CylSP method is accurate than the HO method. For the spherical and triaxial models, the HO method is significantly better than the CylSP method if the particle number is smaller than $5\times10^5$.  However, for the spherical and triaxial Dehnen models, the CylSP method is significantly more accurate than the HO method if the particle number is  $5\times10^6$.                   

For the multicomponent galaxy model with a Ferrers bar, MN disk and the Plummer halo, the HO method has a significant cusp in the central region of the bar, while the CylSP method gives good results for both the analytical model and the $N$-body realization.  
If we consider the bar component of this $N$-body realization separately, the CylSP method can also generate some wiggles in the central region for the force if a small $R_{\rm min}$ is used, but these wiggles vanish with increasing particle numbers. 
The force error  in the central region from the HO method can affect the orbit structure significantly, especially for box orbits, which approach very closely to the center, thus affecting the SOS. We find more chaotic orbits if the HO method is used and also that the stability of the x1 orbits from these two methods may be different. The effect on detailed orbit families and a frequency analysis of  orbits from these two methods will be considered in a future study.

If the input model is an analytical model,  the CylSP method with $R_{\rm min}$ as provided by the AGAMA software is preferred. For models with snapshots of an $N$-body simulation,  the initialization CPU time for the HO method is significantly smaller than that for the CylSP method, while the subsequent CPU time for the CylSP method is much lower than that for the HO method. For models with snapshots of an $N$-body simulation, if the model is a spherical or nearly spherical Hernquist model, the HO method can be preferentially used.

For a multicomponent model for a barred galaxy, as expected, the CylSP method works better. A detailed comparison between these two methods for the particle realization is still needed if a density model other than the Dehnen model studied here is used. If the CylSP method is selected for snapshots of an $N$-body simulation, a relatively large $R_{\rm min}$ ($R_{\rm min}=0.1$) is suggested.

\begin{acknowledgements}
We thank E. Vasiliev and Jean-Charles Lambert  for many useful discussions and for help with software. E. Vasiliev  also gave us insightful comments, which greatly improved the paper.
  The calculations for the CylSP method are made using the publicly available AGAMA software by Vasiliev (2019), and the results of for snapshots of $N$-body simulations for the HO method is performed by the SCF.F software provided by Lars Hernquist. We acknowledge the support by the National Key R\&D Program (No. 2018YFA0404501, 2017YFA0402603), the  National Science Foundation of China (Grant No.11821303, 11773034, 11761141012,  11761131004 and 11633004), the Chinese Academy of Sciences (CAS) Strategic Priority Research Program XDA15020200 and and the CAS Interdisciplinary Innovation Team (JCTD- 2019-05).  

The``Centre de Calcul Intensif d'Aix-Marseille" is acknowledged for granting access to its high performance computing resources, where the initial part of this work was performed. 
Most of the computing was performed on the high performance computing cluster at the Information and Computing Center
at National Astronomical Observatories, Chinese Academy of  Sciences. This work is also supported by Astronomical Big Data Joint Research Center, co-founded by National Astronomical Observatories, Chinese Academy of Sciences and Alibaba Cloud.
\end{acknowledgements}

\bibliographystyle{aa}
\bibliography{ms}

%\begin{appendix}
% \end{appendix}

\end{document}